\documentclass{cernyrep}
\usepackage[T1]{fontenc}
\usepackage[bookmarks=false,colorlinks=true,linkcolor=black, citecolor=black, urlcolor=blue]{hyperref}

\usepackage{latexsym}
\usepackage{amsmath,bm}

\usepackage{multicol,parcolumns}
\usepackage{empheq}

\usepackage{makeidx}
\usepackage{graphicx,url,bm,amsmath,amssymb}
\usepackage{amsfonts}
\usepackage{trfsigns}
\usepackage{units}
\usepackage{booktabs}

\usepackage{fancyhdr}
\fancyhfoffset{4 mm}
\fancypagestyle{ARTTITLE}{% 
\fancyhf{} % clear all header and footer fields
\lhead{\hfill CERN Accelerator School Proceedings ---
{\it RF for Accelerators} ---  Berlin,  Germany, 2023\hfill}
\lfoot{\hspace{3mm} Available online at \url{https://cas.web.cern.ch/previous-schools}}
\rfoot{\thepage\hspace*{3mm}}
 
}

\usepackage{varwidth}
\usepackage{xcolor}
\def\Real{{\rm Re}}

\def\ph#1{{\underline{\hat{#1}}}}

\def\diff{{\rm d}}
\def\tdq#1#2{\frac{{\rm d} #1}{{\rm d} #2}}

\def\rahmen#1{\fbox{$ \displaystyle #1 $}}

\graphicspath{{Fig/}}

\begin{document}
\title{Magnetic Alloy/Ferrite Cavities}

\author{H. Klingbeil}

\institute{Technische Universität Darmstadt, Darmstadt, Germany and \\
GSI Helmholtzzentrum für Schwerionenforschung GmbH, Darmstadt, Germany}

\begin{abstract}
RF cavities loaded with magnetic alloy (MA) or ferrite ring cores are used in synchrotrons and storage
rings if the maximum RF frequency is in the order of a few MHz. A simple model for the description
of cavities of this type is derived. The most important parameters are defined, and some properties
of the material and of the cavity are summarized. Different cavity configurations, development
aspects, and several practical topics are covered. A few specific ferrite- and MA-loaded cavity
systems are discussed as examples.
\end{abstract}


\maketitle
\thispagestyle{ARTTITLE}

\section{Introduction}

\nocite{Brennan1999}

This contribution is based on the article ``Ferrite cavities''~\cite{KlingbeilCAS2010} 
published in the CERN Report CERN-2011-007 
under the CC BY 3.0 Attribution Licence~\cite{CC30} and on  
Section~4.1 of the textbook~\cite{Klingbeil2015} which was re-published in 2022 as an open-access publication under the 
CC BY-NC-ND  licence. 
The original content~\cite{KlingbeilCAS2010link} has been modified and extended.   
 
The revolution frequency of charged particles in synchrotrons or storage rings 
is usually lower than $10 \; \rm MHz$. Even if we consider comparatively small 
synchrotrons (e.g.~like HIT in Heidelberg, Germany, or CNAO in Pavia, Italy, 
with about $20...25 \; \rm m$ diameter, both used for tumor therapy), the 
revolution time will be greater than $200 \;\rm ns$ since the particles cannot
reach the speed of light. Because, according to $f_{\rm RF}=h \cdot f_{\rm R}$,
the RF frequency is an integer multiple of the revolution frequency, the 
RF frequency will typically be lower than $10 \; \rm MHz$ if only small harmonic 
numbers $h$ are desired. 
For such an operating frequency, the spatial dimensions of a conventional 
RF resonator would be far too large to be used in a~synchrotron. 
One possibility to solve this problem is to reduce the wavelength by 
filling the cavity with magnetic material. This is the basic idea of 
magnetic alloy (MA)-loaded cavities or 
ferrite-loaded cavities. Furthermore, this type of cavity offers a simple 
means to modify the resonant frequency in a~wide range 
(typically up to a factor of $10$) and in a comparatively short time 
(typically at least $10 \; \rm ms$ cycle time \cite{Brennan1999}). 
The possibility to adjust the resonant frequency of a
cavity to the desired operating frequency is called \textbf{tuning}. 
An alternative is the construction of an untuned cavity with sufficient bandwidth. 
In any case, MA- or ferrite-loaded cavities are suitable for ramped operation 
in a synchrotron.

\section{Permeability of Magnetic Materials}
\label{ferrites}

In this contribution, all calculations are based on permeability quantities $\mu$ for 
which $\mu=\mu_{\rm r} \mu_0$
holds. In material specifications, the relative permeability $\mu_{\rm r}$ is given
which means that we have to multiply with $\mu_0$ to obtain $\mu$. This comment
is also valid for the incremental/differential permeability introduced in the~following. 

\index{Hysteresis|(}

\index{Magnetic material!soft|see{Soft magnetic material}}
\index{Magnetic material!hard|see{Hard magnetic material}}
\index{Soft magnetic material|(}
\index{Hard magnetic material|(}

In RF cavities, only so-called \textbf{soft magnetic materials} which have a narrow 
hysteresis loop are of interest since their losses are comparatively low
(in contrast to \textbf{hard magnetic materials} which are used for 
permanent magnets\footnote{No strict separation exists between hard and soft magnetic materials.}). 

\index{Soft magnetic material|)}
\index{Hard magnetic material|)}

\begin{figure}[htb]
\begin{minipage}[t]{5.1cm}
%\epsfxsize=6cm
%\epsffile{\bilderpfad/bilder/hysterese2.eps}
\includegraphics[width=5.1cm]{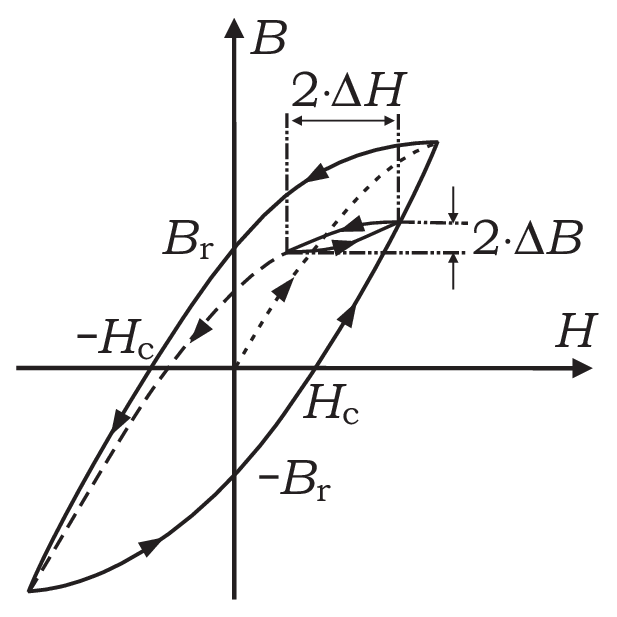}
\caption[bild]{Hysteresis loop.}
\label{hysteresis}
\end{minipage}
\hfill
\begin{minipage}[t]{8.5cm}
%\epsfxsize=10cm
%\epsffile{\bilderpfad/bilder/Kavitaet_3D_d.eps}
\includegraphics[width=8.5cm]{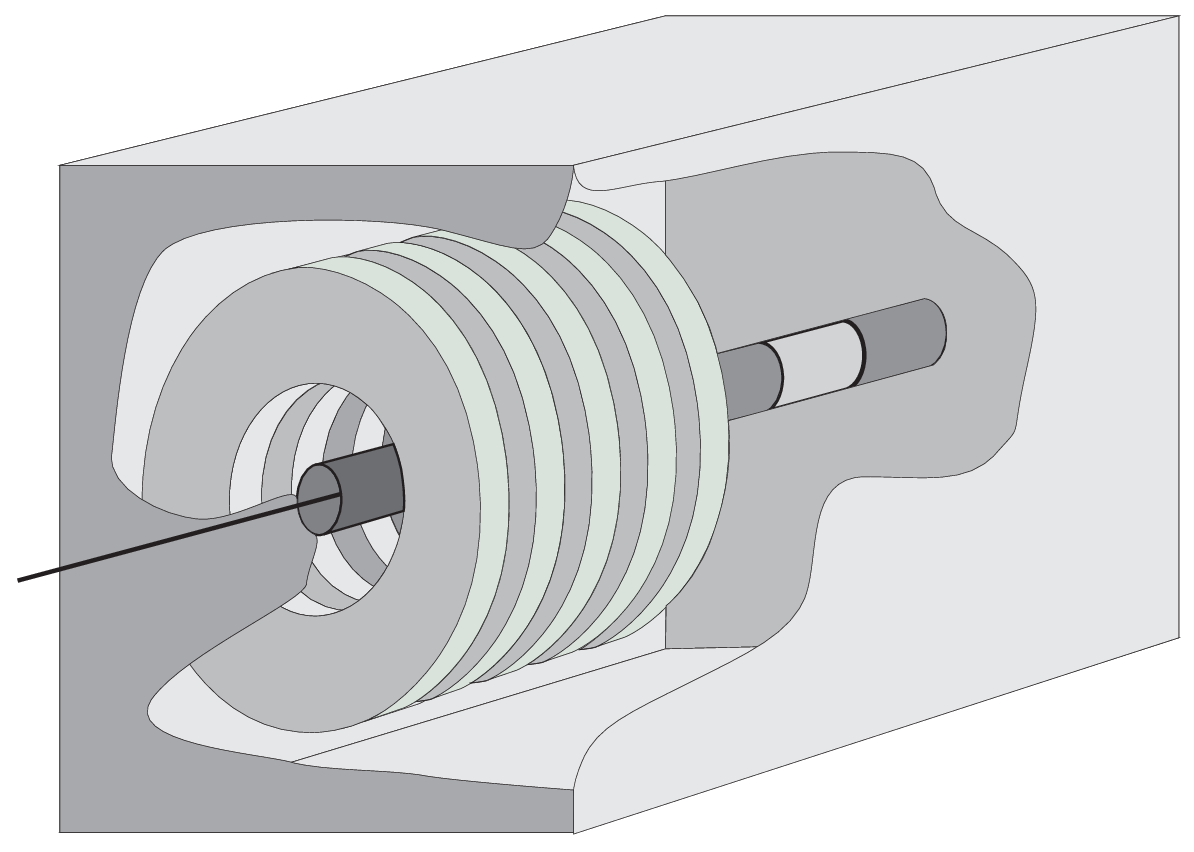}
\caption[bild]{Simplified 3D sketch of a ferrite- or MA-loaded cavity.}
\label{cavity_3d}
\end{minipage}
\end{figure}

Figure~\ref{hysteresis} shows a \textbf{hysteresis loop} of a ferromagnetic material (not necessarily the full loop, but possibly a minor loop -- see e.g. Ref.~\cite{Henze2003}). 
It is well-known that the hysteresis loop leads to a \index{Residual induction}\textbf{residual induction} $B_{\rm r}$ 
if no magnetizing field $H$ is present and that some \index{Coercive magnetizing field}\textbf{coercive magnetizing field}  
$H_{\rm c}$ is needed to set the induction $B$ to zero.\footnote{
The quantities $B_{\rm r}$ and $H_{\rm c}$ are usually defined for the full hysteresis loop with saturation, but they can also be defined for symmetrical minor loops. 
} 

\index{Magnetizing field!coercive|see{Coercive magnetizing field}}
\index{Induction!residual|see{Residual induction}}
\index{Bias current|see{Biasing}}
\index{Incremental permeability|(}
\index{Differential permeability|(}
\index{Biasing|(}

Let us now assume that some cycles of the larger hysteresis loop in Fig.~\ref{hysteresis} have already 
passed and that $H$ is currently increasing. 
We now stop to increase the magnetizing field $H$ in the upper right 
part of the diagram. Then, $H$ is decreased by a much smaller amount 
$2 \cdot \Delta H$, afterwards increased again by that amount $2 \cdot \Delta H$, 
and so forth\footnote{
The factor of $2$ was assumed in order to have the same total change 
of $2 \cdot \Delta H$ as in the equation 
\[
H_{\rm AC}(t)=\Delta H \; \cos(\omega t) 
\]
which is usually used for harmonic oscillations. 
}. 
As the 
diagram shows, this procedure will lead to a much smaller hysteresis loop 
where $B$ changes by $2 \cdot \Delta B$. We may therefore define 
a \textbf{differential} or \textbf{incremental permeability}\footnote{
In a strict sense, the differential permeability is the limit 
\[
\mu_\Delta=\tdq{B}{H}
\]
for $\Delta H, \Delta B \rightarrow 0$.
} 
\[
\rahmen{
\mu_\Delta=\frac{\Delta B}{\Delta H}
}
\]
which describes the slope of the local hysteresis loop. It is this quantity 
$\mu_\Delta$ which is relevant for RF applications. One can see that  
$\mu_\Delta$ can be decreased by increasing the DC component of $H$
thereby driving the material closer to saturation. 
Since $H$ is generated by currents, one speaks of a \textbf{bias current} that is 
applied in order to shift the \index{Operating point}\textbf{operating point} to 
higher inductions $B$ 
leading to a lower differential permeability $\mu_\Delta$. 

\index{Incremental permeability|)}
\index{Differential permeability|)}

\index{Loss!magnetic|see{Magnetic loss}}
\index{Loss!hysteresis|see{Hysteresis loss}}
\index{Loss!eddy current|see{Eddy current loss}}
\index{Loss!residual|see{Residual loss}}

\index{Complex permeability|(}
\index{Preisach model|(}
\index{Magnetic loss|(}
\index{Hysteresis loss|(}
\index{Eddy current loss|(}
\index{Residual loss|(}

The hysteresis loops and the AC~permeability of ferromagnetic materials 
can be described in a~phenomenological way by the so-called \textbf{Preisach model}
(Ref.~\cite{Mayergoyz}) or similar models~\cite{Takach1995}. 
Unfortunately, the material properties are even more complicated since 
they are also frequency-dependent. One usually uses the 
\textbf{complex permeability} 
\begin{equation}
\rahmen{
\underline \mu=\mu'_{\rm s} -j \mu''_{\rm s}
}
\label{glpermeab1}
\end{equation}
to describe losses 
(hysteresis loss, eddy current loss and residual loss). 
The parameters $\mu'_{\rm s}$ and $\mu''_{\rm s}$ are
frequency-dependent. In the following, we will assume that the complex 
permeability $\underline \mu$ describes the~material behavior in rapidly 
alternating fields as the above-mentioned real quantity $\mu_\Delta$ does
when a~biasing field $H_{\rm bias}$ is present. 
However, we will omit the index $\Delta$ for the sake of simplicity. 

\index{Hysteresis loss|)}
\index{Eddy current loss|)}
\index{Residual loss|)}
\index{Magnetic loss|)}

\index{Preisach model|)}
\index{Biasing|)}
\index{Complex permeability|)}
\index{Hysteresis|)}

\section{Magneto-Quasistatic Analysis of an MA/Ferrite Cavity}
\label{analysis}

In this section, 
we will see that a ferrite- or MA-loaded cavity may roughly be regarded as a 
transformer whose primary coil consists of only one winding  
fed by an RF power source and where the beam interaction with the cavity can be described by   
a secondary coil. Consequently, some conclusions that are valid for 
transformers are also valid here. For example, the cavity will not work properly
if the frequency is too low because the reactance 
(product of inductance and angular frequency) will be too small in comparison 
with the ohmic parts thereby decreasing the transformation ratio. 
If the frequency becomes too large flux leakage and 
distributed effects will become important, so that a simple magneto-quasistatic 
analysis is no longer possible. Hence, an optimum operating frequency range can 
be specified for an MA- or ferrite-loaded cavity similar to a transformer taking the
material properties and the geometry into account. 
It is assumed without further notice that the considered frequency belongs 
to this optimum frequency range.

\index{Gap!ceramic|see{Ceramic gap}}
\index{Ceramic gap|(}

Figure~\ref{cavity_3d} shows the main elements of a ferrite-loaded cavity. 
The conductive beam pipe is interrupted by a \textbf{ceramic gap}. This 
gap ensures that the beam pipe may still be evacuated but it allows a~voltage $V_{\rm gap}$ to be induced in longitudinal direction. Several MA or ferrite
\index{Ring core}\textbf{ring cores} are mounted in a~concentric way around beam and beam pipe
(five ring cores are drawn here as an example).
The whole cavity is surrounded by a metallic housing which is connected to
the beam pipe.

\index{Ceramic gap|)}

\begin{figure}[ht]
\begin{center}
\begin{minipage}[t]{10.2cm}
%\epsfxsize=12cm
%\epsffile{\bilderpfad/bilder/cavity2.eps}
\includegraphics[width=10.2cm]{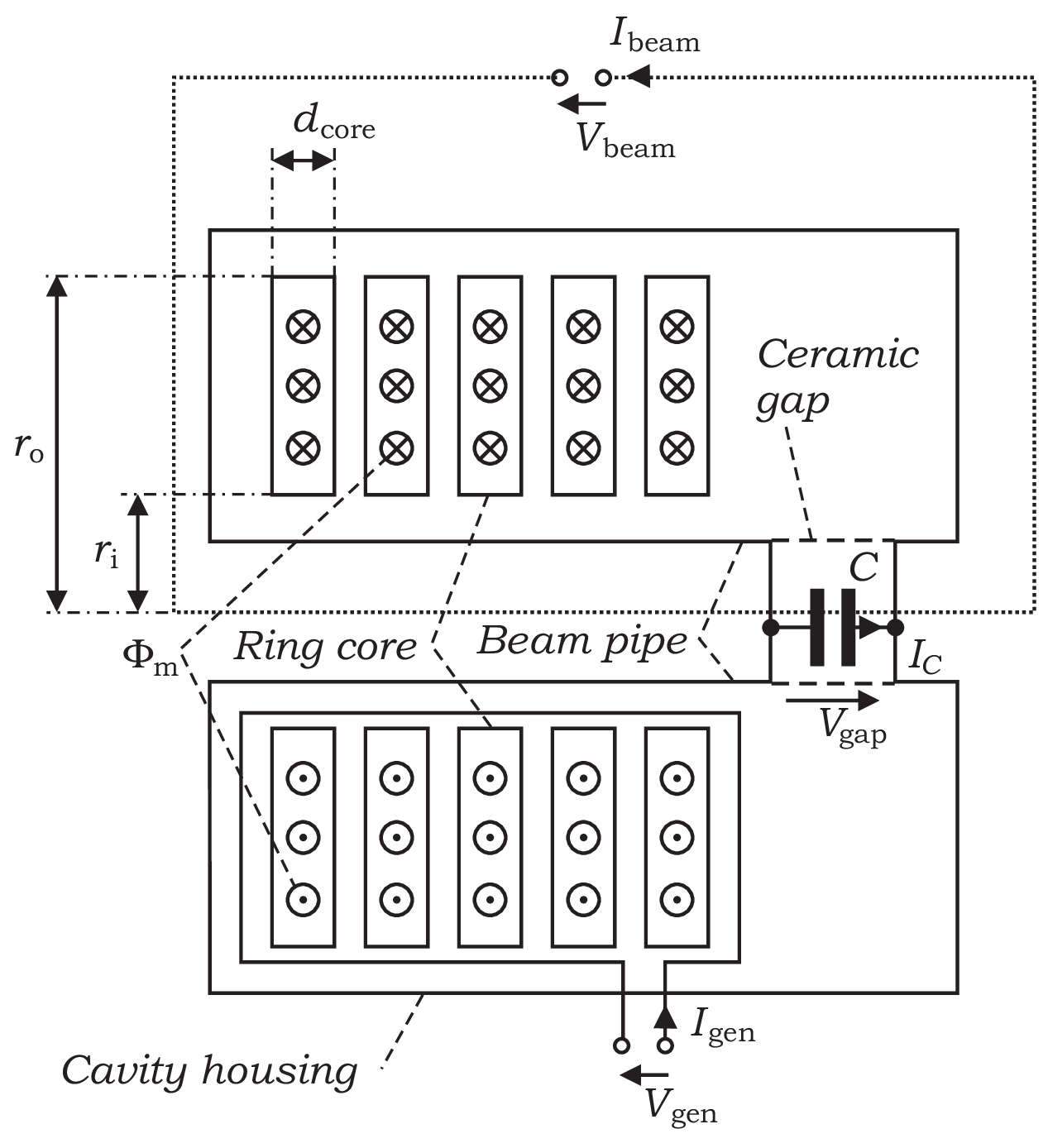}
\end{minipage}
\end{center}
\caption[bild]{Simplified wire model of a ferrite or MA cavity.}
\label{cavity}
\end{figure}

Figure~\ref{cavity} shows a cross-section through the cavity.
The dotted line represents the beam which is located 
in the middle of the metallic beam pipe (for analyzing the influence of the
beam current, this dotted line is regarded as a part of a circuit that closes 
outside the cavity, but this is not relevant for understanding the basic 
operation principle). The ceramic gap has a parasitic capacitance, further distributed capacitances are always present, and additional lumped-element capacitors are
often connected in parallel to the gap ---~leading to the overall capacitance $C$.
Starting at the generator port located at the~bottom of the figure, an 
inductive \index{Coupling loop}coupling loop surrounds the ring core stack. 
This loop was not shown in Fig.~\ref{cavity_3d}.

Due to the cross-section approach, we get a wire model of the 
cavity with ideal wires representing the cavity housing. This is sufficient for 
the practical analysis, but one should remember that the currents are 
distributed in reality.  

\index{Phasor|(}

In the following, 
voltages, currents, field and flux quantities will be represented by phasors, 
i.e.~complex amplitudes/peak values for a given frequency 
$f=\omega/2\pi$. For a quantity $X$ in time domain, 
we write $\ph{X}$ for the phasor in frequency domain. The function $X(t)$ can
be reconstructed by means of the complex function 
$\underline{X}=\ph{X} e^{j\omega t}$ 
according to
$X(t)=\Real\{\underline{X}\}=\Real\{\ph{X} e^{j\omega t}\}$.

\index{Phasor|)}

\begin{figure}[htb]
\begin{minipage}[t]{6.4cm}
%\epsfxsize=7.5cm
%\epsffile{\bilderpfad/bilder/ESB_RLC_Serial1.eps}
\includegraphics[width=6.4cm]{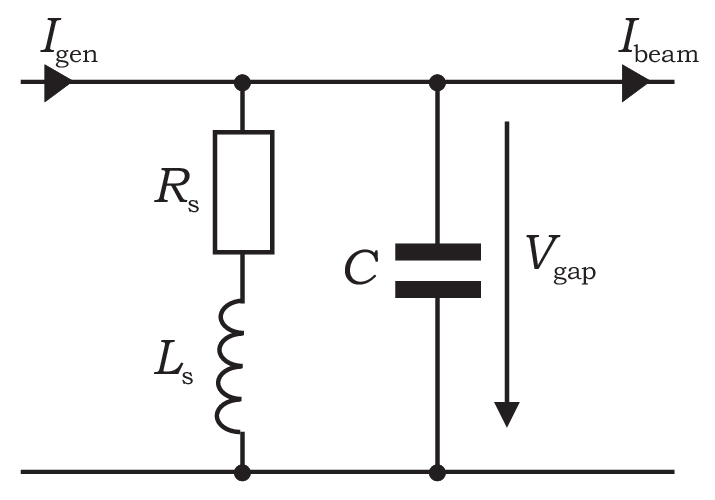}
\caption[bild]{Series equivalent circuit.}
\label{esbrlc1}
\end{minipage}
\hfill
\begin{minipage}[t]{6.4cm}
%\epsfxsize=7.5cm
%\epsffile{\bilderpfad/bilder/ESB_RLC_Parallel1.eps}
\includegraphics[width=6.4cm]{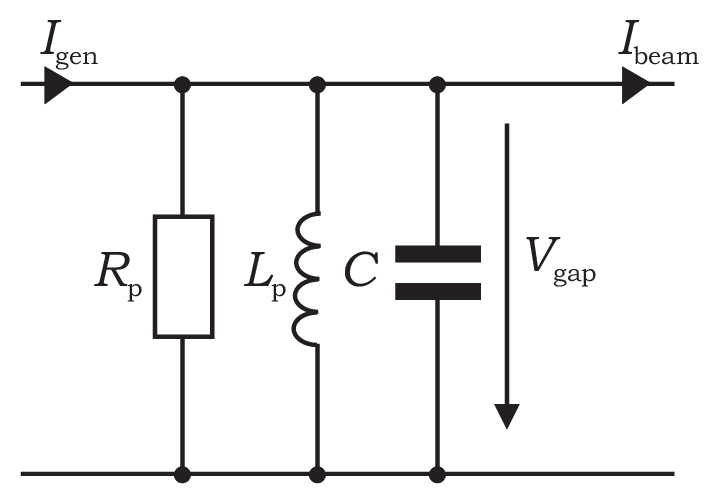}
\caption[bild]{Parallel equivalent circuit.}
\label{esbrlc2}
\end{minipage}
\end{figure}

Let us consider the coupling loop contour in Fig.~\ref{cavity} which surrounds the lower left ring core stack cross-section. 
Based on Maxwell's second equation in the time domain (Faraday's law) 
\[
\oint_{\partial A} \vec E \cdot \diff \vec r=-\int_A \dot{\vec B} \cdot \diff \vec A 
\] 
we find  
\begin{align*}
\ph{V}_{\rm gen}=+j \omega \ph{\Phi}_{\rm m,tot}
%\label{glua1}
\end{align*}
as the dependence of the generator voltage on the magnetic flux in the frequency domain. 
If we now use the complete lower cavity half or the beam current contour as integration paths, one obtains  
\begin{align}
\ph{V}_{\rm gap}=+j \omega \ph{\Phi}_{\rm m,tot}, \mbox{\quad}
\ph{V}_{\rm beam}=+j \omega \ph{\Phi}_{\rm m,tot} \mbox{\quad}
\Rightarrow 
\ph{V}_{\rm beam}=\ph{V}_{\rm gap} =\ph{V}_{\rm gen} \label{glug1}.
\end{align}
Here we assumed that the stray field $B$ in the air regions is negligible in 
comparison with the field inside the ring cores (due to their high permeability).
For negligible displacement current we have Maxwell's first equation (Amp\`ere's law)
\[
\oint_{\partial A} \vec H \cdot \diff \vec r=\int_A \vec J \cdot \diff \vec A. 
\] 
We use a concentric circle with radius $\rho$ around the beam axis as integration path: 
\begin{align*}
H \; 2 \pi \rho=I_{\rm tot}.
%\label{gliges1}
\end{align*}
In the frequency domain, this leads to 
\begin{equation}
\ph{B}=\underline \mu \frac{\ph{I}_{\rm tot}}{2 \pi \rho}
\label{glbfield1}
\end{equation}
with  
\begin{equation}
\ph{I}_{\rm tot}=\ph{I}_{\rm gen}-\ph{I}_C-\ph{I}_{\rm beam}.
\label{gliges2}
\end{equation}
For the flux through one single ring core we get  
\[
\ph{\Phi}_{\rm m,1}=\int \ph{\vec B} \cdot \diff \vec A=d_{\rm core} \; \int_{r_{\rm i}}^{r_{\rm o}} \ph{B} \; \diff \rho=
\frac{d_{\rm core} \underline \mu \ph{I}_{\rm tot}}{2 \pi} \; \ln \frac{r_{\rm o}}{r_{\rm i}}.
\]
With the complex permeability $\underline \mu=\mu'_{\rm s}-j \mu''_{\rm s}$ 
and assuming that $N$ ring cores are present, Eq.~(\ref{glug1}) provides 
\[
\ph{V}_{\rm gap}=j \omega \ph{\Phi}_{\rm m,tot}=j \omega N \; \ph{\Phi}_{\rm m,1}=
j \omega \frac{N d_{\rm core} (\mu'_{\rm s}-j \mu''_{\rm s}) \ph{I}_{\rm tot}}{2 \pi} \; \ln \frac{r_{\rm o}}{r_{\rm i}}.
\]  
Therefore, we obtain 
\begin{equation}
\ph{V}_{\rm gap}=\ph{I}_{\rm tot} (j \omega L_{\rm s}+R_{\rm s})=\ph{I}_{\rm tot} Z_{\rm s},
\label{glua2}
\end{equation}
if 
\begin{equation}
Z_{\rm s}=\frac{1}{Y_{\rm s}}=j \omega L_{\rm s}+R_{\rm s},
\label{gldefz}
\end{equation}
\begin{align}
\rahmen{
L_{\rm s}=\frac{N d_{\rm core} \mu'_{\rm s}}{2 \pi} \; \ln \frac{r_{\rm o}}{r_{\rm i}}
} , \mbox{\quad}
\rahmen{
R_{\rm s}=\omega \frac{N d_{\rm core} \mu''_{\rm s}}{2 \pi} \; \ln \frac{r_{\rm o}}{r_{\rm i}}
=\omega \frac{\mu''_{\rm s}}{\mu'_{\rm s}} L_{\rm s}=\frac{\omega L_{\rm s}}{Q}
} 
\label{glqfactor1}
\end{align}
are defined. 
Here, 
\begin{align}
Q=\frac{\mu'_{\rm s}}{\mu''_{\rm s}}=\frac{1}{\tan \delta_\mu}
\label{glqfactor2}
\end{align}
is the \index{Q factor|(} \textbf{quality factor} (or \textbf{Q factor}) of the ring core material. 
Using Eq.~(\ref{gliges2}) we find: 
\begin{align}
\ph{V}_{\rm gap} Y_{\rm s}=\ph{I}_{\rm tot} =\ph{I}_{\rm gen}-\ph{I}_{\rm beam}-\ph{V}_{\rm gap} \; j \omega C
\Rightarrow
\rahmen{
\ph{V}_{\rm gap}=\frac{\ph{I}_{\rm gen}-\ph{I}_{\rm beam}}{Y_{\rm s}+j\omega C}=Z_{\rm tot}(\ph{I}_{\rm gen}-\ph{I}_{\rm beam})
}
\label{glcavimpedance1}
\end{align}
This equation corresponds to 
the \index{Cavity!equivalent circuit|(} equivalent circuit 
shown in Fig.~\ref{esbrlc1}.
In the last step, we defined $Y_{\rm tot}=1/Z_{\rm tot}=Y_{\rm s}+j\omega C$.
In literature one often finds a different version of Eq.~(\ref{glcavimpedance1}) where $\ph{I}_{\rm beam}$ has the same sign as $\ph{I}_{\rm gen}$.
This corresponds to both currents having the~same direction (flowing 
into the circuits in Figs.~\ref{esbrlc1} and \ref{esbrlc2}). In any case, 
one has to make sure that the correct phase between beam current and 
gap voltage is established.  

\index{Impedance!beam|see{Beam impedance}}
\index{Beam loading|(}
\index{Beam impedance|(}

Assuming that the gap voltage is given by $V_{\rm gap}(t)=\hat V_{\rm gap} \sin(\omega_{\rm RF} t)$, 
the bunches will be located at $t=0,\pm T_{\rm RF},\pm 2 T_{\rm RF},\dots$ if the 
stationary case without acceleration is considered 
(operation with positively charged particles below transition energy).
Therefore, the fundamental harmonic of the beam current will be proportional 
to $\cos(\omega_{\rm RF} t)$ which corresponds to a $90^\circ$ phase shift 
between $V_{\rm gap}$ and $I_{\rm beam}$. For low beam currents and for a 
cavity that is tuned to resonance, the phase of the gap voltage is equal to 
the phase of the generator current. For higher beam currents, however, not only
the generator current, but also the beam current will have an influence on the 
gap voltage due to the \textbf{beam impedance} $Z_{\rm tot}$
as one sees in Eq.~(\ref{glcavimpedance1}) and in Fig.~\ref{esbrlc1}. 
This phenomenon is called \textbf{beam loading}.  

\index{Beam impedance|)}
\index{Beam loading|)}

\index{Equivalent circuit!of cavity|see{Cavity, equivalent circuit}}
\index{Lumped element circuit!of cavity|see{Cavity, equivalent circuit}}

\section{Parallel and Series Lumped Element Circuit}

Assuming frequency-dependent values\footnote{Even if all circuit elements are assumed to be frequency-independent, both equivalent circuits lead to a similar impedance curve in the vicinity of the resonance frequency.} $R_{\rm s}$, $L_{\rm s}$, $R_{\rm p}$, $L_{\rm p}$, it is possible to convert the 
lumped element circuit shown in Fig.~\ref{esbrlc1} into a parallel circuit 
as shown in Fig.~\ref{esbrlc2}. The admittance of both circuits shall be 
equal: 
\begin{align*}
Y_{\rm tot}=j \omega C + \frac{1}{R_{\rm s}+j \omega L_{\rm s}}=j \omega C +\frac{1}{R_{\rm p}}+\frac{1}{j \omega L_{\rm p}}
\mbox{\quad}
\Rightarrow 
\frac{R_{\rm s}-j \omega L_{\rm s}}{R_{\rm s}^2+(\omega L_{\rm s})^2}=\frac{1}{R_{\rm p}}+\frac{1}{j \omega L_{\rm p}}
\end{align*}
A comparison of real and imaginary part yields 
\begin{align} 
\rahmen{
R_{\rm p} = \frac{R_{\rm s}^2+(\omega L_{\rm s})^2}{R_{\rm s}} 
}, 
\rahmen{
\omega L_{\rm p} = \frac{R_{\rm s}^2+(\omega L_{\rm s})^2}{\omega L_{\rm s}}  
}\label{gllp1}
%\label{glrp1} 
\end{align}
\begin{align}
\Rightarrow R_{\rm p} R_{\rm s}=(\omega L_{\rm p}) (\omega L_{\rm s}).
\label{glrplp1}
\end{align}
For the inverse relation, we write the first Eq.~(\ref{gllp1}) in the form $(\omega L_{\rm s})^2=R_{\rm s}(R_{\rm p} -R_{\rm s})$
and use this result in Eq.~(\ref{glrplp1}):  
\begin{align}
\omega L_{\rm p} \sqrt{R_{\rm s}(R_{\rm p} -R_{\rm s})}=R_{\rm s} R_{\rm p}
\mbox{\quad}
\Rightarrow
(\omega L_{\rm p})^2 (R_{\rm p} -R_{\rm s})=R_{\rm s} R_{\rm p}^2
\mbox{\quad}
\Rightarrow
\rahmen{
R_{\rm s}=\frac{(\omega L_{\rm p})^2}{R_{\rm p}^2+(\omega L_{\rm p})^2} R_{\rm p}
}
\end{align}
With this formula, Eq.~(\ref{glrplp1}) leads to 
\[
\rahmen{
\omega L_{\rm s}=\frac{R_{\rm p}}{\omega L_{\rm p}} R_{\rm s}=\frac{R_{\rm p}^2}{R_{\rm p}^2+(\omega L_{\rm p})^2} \omega L_{\rm p}
}.
\]
Since it is suitable to use both types of lumped element circuit, it is 
also convenient to define 
the \index{Complex permeability}complex permeability $\underline \mu$ in 
a parallel form: 
\begin{equation}
\rahmen{
\frac{1}{\underline \mu}= \frac{1}{\mu'_{\rm p}} +j \frac{1}{\mu''_{\rm p}}
}.
\label{glpermeab2}
\end{equation} 
This is an alternative representation for the series form shown in Eq.~(\ref{glpermeab1}) which leads to 
\[
\frac{1}{\underline \mu}=\frac{\mu'_{\rm s} +j \mu''_{\rm s}}{{\mu'_{\rm s}}^2+{\mu''_{\rm s}}^2}.
\]
Comparing the real and imaginary parts of the last two equations, we find 
\begin{align}
\mu'_{\rm p}=\frac{{\mu'_{\rm s}}^2+{\mu''_{\rm s}}^2}{\mu'_{\rm s}}, 
\mbox{\quad}
\mu''_{\rm p}=\frac{{\mu'_{\rm s}}^2+{\mu''_{\rm s}}^2}{\mu''_{\rm s}}
\mbox{\quad}
\Rightarrow 
\mu'_{\rm p} \mu'_{\rm s}=\mu''_{\rm p} \mu''_{\rm s}.
\label{glmu1}
\end{align}
Together with Eqs.~(\ref{glqfactor1}), (\ref{glqfactor2}),
and (\ref{glrplp1}) we conclude: 
\begin{equation}
\rahmen{
Q=\frac{\mu'_{\rm s}}{\mu''_{\rm s}}=\frac{\omega L_{\rm s}}{R_{\rm s}}
=\frac{R_{\rm p}}{\omega L_{\rm p}}=\frac{\mu''_{\rm p}}{\mu'_{\rm p}}
}.
\label{glqfactor3}
\end{equation}
With these expressions, we may write Eq.~(\ref{glmu1}) in the form
\begin{align}
\rahmen{
\mu'_{\rm p}=\mu'_{\rm s} \left(1 +\frac{1}{Q^2}\right)
}, \mbox{\quad}
\rahmen{
\mu''_{\rm p}=\mu''_{\rm s} \left(1 +Q^2 \right)
}.
\label{glmu3}
%\label{glmu4}
\end{align}
If we use $Q=\omega L_{\rm s}/R_{\rm s}$ from Eq.~(\ref{glqfactor3}), we may rewrite Eq.~(\ref{gllp1}) in the form 
\begin{align} 
\rahmen{
R_{\rm p} = R_{\rm s} (1+Q^2) 
}, \mbox{\quad}
\rahmen{
L_{\rm p} = L_{\rm s} \left(1+\frac{1}{Q^2} \right)
}. 
%\label{gllp2}
\label{glrp2} 
\end{align}
By combining Eqs.~(\ref{glrp2}) and (\ref{glqfactor1}) we find
\[
R_{\rm p}=(1+Q^2)\omega \frac{N d_{\rm core} \mu''_{\rm s}}{2 \pi} \; \ln \frac{r_{\rm o}}{r_{\rm i}}.
\]
With the help of Eqs.~(\ref{glqfactor3}) and (\ref{glmu3}) one gets 
\[
R_{\rm p} = \omega \frac{N d_{\rm core} \mu''_{\rm p}}{2 \pi} \; \ln \frac{r_{\rm o}}{r_{\rm i}}
= \omega \frac{N d_{\rm core} \mu'_{\rm p} Q}{2 \pi} \; \ln \frac{r_{\rm o}}{r_{\rm i}}
=N d_{\rm core} \mu'_{\rm p} Q f \; \ln \frac{r_{\rm o}}{r_{\rm i}}.
\]
This shows that $R_{\rm p}$ is proportional to the product $\mu'_{\rm p} Q f$ which is
a material property. The other parameters refer to the geometry.  
Therefore, the manufacturers of ferrite cores or MA cores sometimes specify 
the \index{mu Q f product@$\mu_{\rm r} Q f$ product|(} {$\bf \mu_{\rm r} Q f$ \bf product} 
(for the sake of simplicity,
we define\footnote{
Here, the index $r$ again denotes the relative permeability, i.e.
$\mu'_{\rm p,r}=\mu'_{\rm p}/\mu_0$ 
} $\mu_{\rm r}:=\mu'_{\rm p,r}$).  

\index{mu Q f product@$\mu_{\rm r} Q f$ product|)}
\index{Permeability!complex|see{Complex permeability}}

For $Q \ge 5$ (typically fulfilled for ferrites), we may use the approximations 
\begin{equation}
R_{\rm p} \approx R_{\rm s} \; Q^2, \mbox{\qquad}
L_{\rm p} \approx L_{\rm s}, \mbox{\qquad}
\mu'_{\rm p} \approx \mu'_{\rm s}, \mbox{\qquad}
\mu''_{\rm p} \approx \mu''_{\rm s} \; Q^2
\label{glapproxferrite1}
\end{equation}
which then have an error of less than $4 \; \%$. 

\index{Q factor|)} 
\index{Quality factor|see{Q factor}}

\index{Cavity!equivalent circuit|)}

\section{Frequency Dependence of Material Properties}
\label{material}

\begin{figure}[htb]
\begin{minipage}{7.5cm}
\centering
%\includegraphics[height=6cm]{\bilderpfad/bilder/Ferroxcube_Re_mu.eps}
%\epsfxsize=7.5cm
%\epsffile{\bilderpfad/bilder/Ferroxcube_Re_mu.eps}
\includegraphics[width=7.5cm]{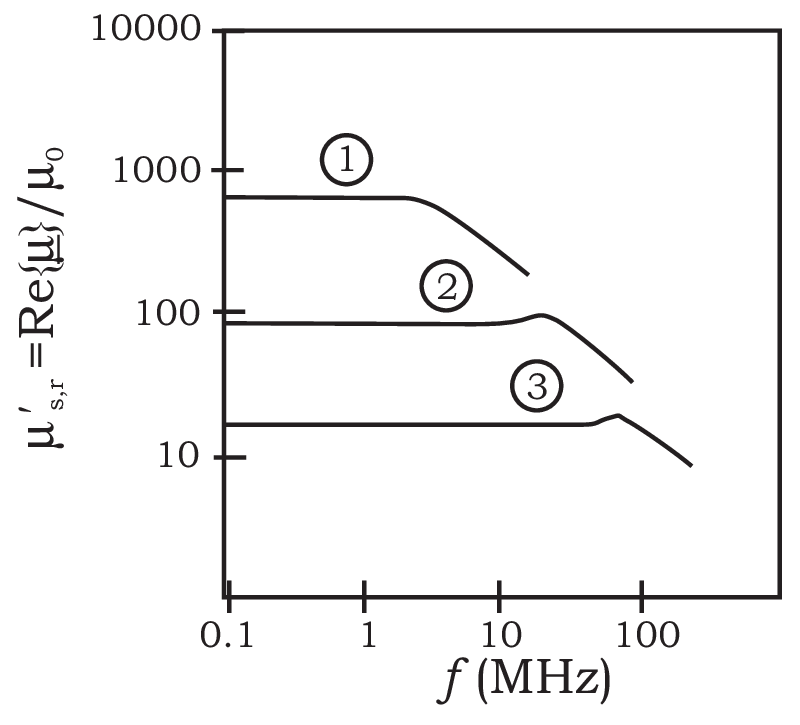}
\caption[bild]{$\mu'_{\rm s,r}$ versus frequency for three different types of 
ferrite material (1: Ferroxcube 4A, 2: Ferroxcube 4C, 3: Ferroxcube 4E). 
Data adopted from \cite{Philips1969}.}
\label{ferroxcube1}
\end{minipage}
\hfill
\begin{minipage}{7.5cm}
\centering
%\includegraphics[height=6cm]{\bilderpfad/bilder/Ferroxcube_Im_mu.eps}
%\epsfxsize=7.5cm
%\epsffile{\bilderpfad/bilder/Ferroxcube_Im_mu.eps}
\includegraphics[width=7.5cm]{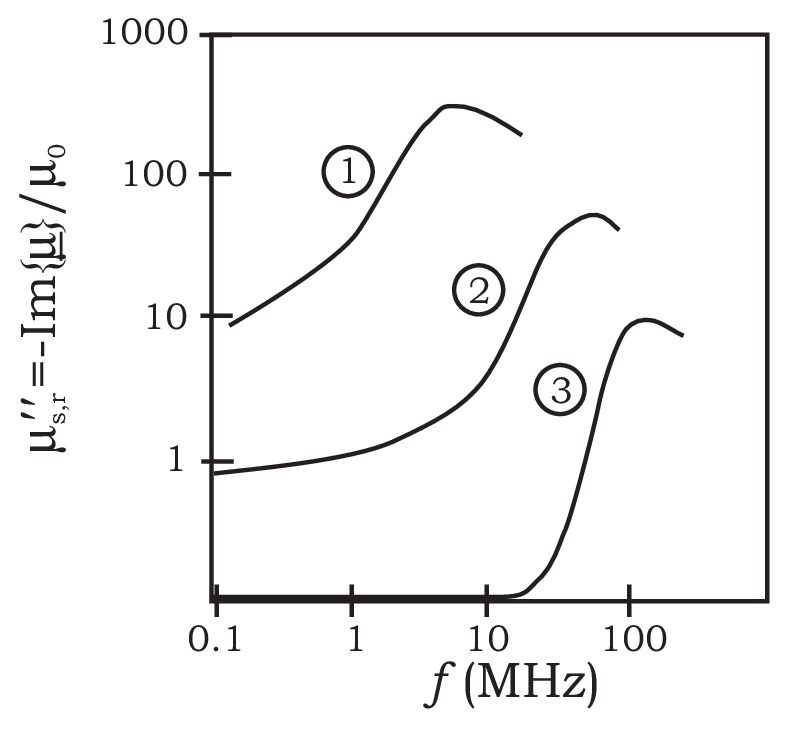}
\caption[bild]{$\mu''_{\rm s,r}$ versus frequency for three different types of 
ferrite material (1: Ferroxcube 4A, 2: Ferroxcube 4C, 3: Ferroxcube 4E).
Data adopted from \cite{Philips1969}.}
\label{ferroxcube2}
\end{minipage}
\end{figure}

\begin{figure}[htb]
\centering
%\includegraphics[height=6cm]{\bilderpfad/bilder/Ferroxcube_muqf1}
%\epsfxsize=7.5cm
%\epsffile{\bilderpfad/bilder/Ferroxcube_muqf1.eps}
\includegraphics[width=7.5cm]{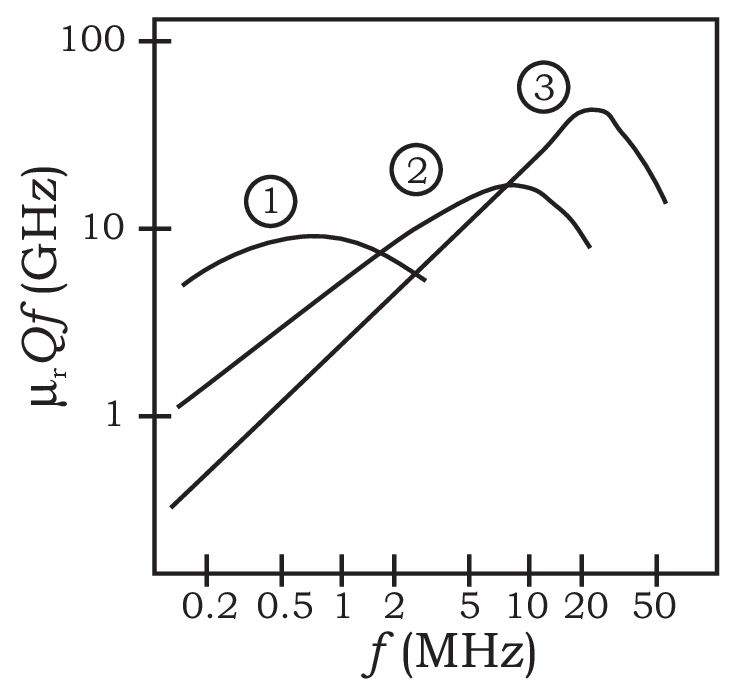}
\caption[bild]{$\mu'_{\rm s,r}Qf$ product versus frequency for three different types of 
ferrite material (1:~Ferroxcube 4A, 2:~Ferroxcube 4C, 3:~Ferroxcube 4E). 
Please note that $\mu_{\rm r}:=\mu'_{\rm p,r} \approx \mu'_{\rm s,r}$ holds due to 
Eq.~(\ref{glapproxferrite1}).
No bias field is present, and small magnetic RF field amplitudes are assumed.
Data adopted from \cite{Philips1969}.}
\label{ferroxcube3}
\end{figure}

As an example, the frequency dependence of the permeability is shown 
in Figs.~\ref{ferroxcube1} and \ref{ferroxcube2} 
for the special ferrite material Ferroxcube 4 assuming small magnetic RF fields
without biasing. 
All the data presented for this material are taken from \cite{Philips1969}.
It is obvious that the 
behavior depends significantly on the choice of the material. 
Without biasing, a constant $\mu'_{\rm s} \approx \mu'_{\rm p}$ may only be assumed up to
a certain frequency (see Fig.~\ref{ferroxcube1}). 
Increasing the frequency from $0$ upwards, the Q factor will decrease
(compare Figs.~\ref{ferroxcube1} and \ref{ferroxcube2}). 
Figure~\ref{ferroxcube3} shows the resulting frequency dependence of the 
$\mu_{\rm r} Qf$ product. 

If the magnetic RF field is increased, both $Q$ and $\mu_{\rm r} Qf$ will 
decrease in comparison with the~diagrams in 
Figs.~\ref{ferroxcube1}-\ref{ferroxcube3}. The effective 
incremental permeability $\mu_{\rm r}$ will increase for rising magnetic RF fields 
as one can see by interpreting Fig.~\ref{hysteresis}. 
Therefore, it is important to consider the material properties under 
realistic operating conditions (the maximum RF B-field for ferrites is usually in the order of
$10 \dots 20 \; \rm mT$).

If biasing is applied, the $\mu_{\rm r} Qf$ curve shown in Fig.~\ref{ferroxcube3}
will be shifted to the lower right side; this effect may approximately compensate
the increase of $\mu_{\rm r} Qf$ with frequency \cite{Philips1969}. 
Therefore, the $\mu_{\rm r} Qf$ product 
may sometimes approximately be regarded as a constant if biasing is used to 
keep the cavity at resonance for all frequencies under consideration.

\section{Quality Factor of the Cavity}

\index{Q factor!of a cavity|see{Cavity, Q factor}}
\index{Q factor|(}
\index{Resonant frequency|(}
\index{Cavity!Q factor|(}

The quality factor of the equivalent circuit shown in Fig.~\ref{esbrlc2} 
is obtained if the \textbf{resonant (angular) frequency}
\[
\rahmen{
\omega_{\rm res}=2\pi f_{\rm res}=\frac{1}{\sqrt{L_{\rm p} C}}
}
\]
is inserted into Eq.~(\ref{glqfactor3}): 
\[
\rahmen{
Q_{\rm p}=R_{\rm p} \sqrt{\frac{C}{L_{\rm p}}}
}.
\]
In general, all parameters $\mu_s'$, $\mu_s''$, $\mu_p'$, $\mu_p''$, 
$R_{\rm s}$, $L_{\rm s}$, $R_{\rm p}$, $L_{\rm p}$, $Q$ and $Q_{\rm p}$ are frequency-dependent.
It depends on the material whether the parallel or the series lumped element 
circuit is the better representation in the sense that its parameters may 
be regarded as approximately constant in the relevant operating range. 
In the following, we will use the parallel representation. 

We briefly show that $Q_{\rm p}$ is in fact the quality factor defined by
$Q_{\rm p}=\omega \overline{W}_{\rm tot}/\overline{P}_{\rm loss}$,
where $\overline{W}_{\rm tot}$ is the stored energy 
and $\overline{P}_{\rm loss}$ is the power loss
(both time-averaged): 
\begin{equation}
\rahmen{
\overline{P}_{\rm loss}=\frac{|\ph{V}_{\rm gap}|^2}{2R_{\rm p}}
}
\label{glpowerloss1}
\end{equation}
\begin{align*}
\overline{W}_{\rm el}=\frac{1}{4} \; C \; |\ph{V}_{\rm gap}|^2, \mbox{\qquad}
\overline{W}_{\rm magn}=\frac{1}{4} \; L_{\rm p} \; |\ph{I}_L|^2
=\frac{1}{4} \; L_{\rm p} \; \frac{|\ph{V}_{\rm gap}|^2}{\omega^2 L_{\rm p}^2}
=\frac{|\ph{V}_{\rm gap}|^2}{4 \omega^2 L_{\rm p}}.
\end{align*}
At resonance, we have $\overline{W}_{\rm el}=\overline{W}_{\rm magn}$ which leads to
\[
Q_{\rm p}=2 \omega \frac{\overline{W}_{\rm el}}{\overline{P}_{\rm loss}}=
2 \omega \frac{R_{\rm p} C}{2}=R_{\rm p} \sqrt{\frac{C}{L_{\rm p}}}
\]
as expected. 
The parallel resistor $R_{\rm p}$ defined by Eq.~(\ref{glpowerloss1}) is sometimes
called \index{Shunt impedance|(} \textbf{shunt impedance}.
Please note that different definitions for the shunt impedance exist in literature. 
Sometimes, especially in the~LINAC community, the shunt impedance is defined 
as two times $R_{\rm p}$ (see Ref.~\cite{Gerigk2011}).

\index{Cavity!Q factor|)}
\index{Shunt impedance|)}
\index{Cavity!shunt impedance|see{Shunt impedance}}

\section{Impedance of the Cavity}

\index{Cavity!impedance|(}

According to Fig.~\ref{esbrlc2}, the impedance of the cavity  
\[
Z_{\rm tot}=\frac{1}{\frac{1}{R_{\rm p}}+j \left(\omega C-\frac{1}{\omega L_{\rm p}} \right)}
=\frac{\sqrt{\frac{L_{\rm p}}{C}}}{\frac{1}{R_{\rm p}}\sqrt{\frac{L_{\rm p}}{C}}+j
\left(\omega \sqrt{L_{\rm p} C}-\frac{1}{\omega \sqrt{L_{\rm p} C}} \right)}
\]
may be written as  
\begin{equation}
Z_{\rm tot}=\frac{\frac{R_{\rm p}}{Q_{\rm p}}}{\frac{1}{Q_{\rm p}}+j
\left(\frac{\omega}{\omega_{\rm res}}-\frac{\omega_{\rm res}}{\omega} \right)}
\mbox{\quad}
\Rightarrow 
\rahmen{
Z_{\rm tot}=\frac{R_{\rm p}}{1+j \; Q_{\rm p}
\left(\frac{\omega}{\omega_{\rm res}}-\frac{\omega_{\rm res}}{\omega} \right)}
}.
\label{glimpedance1}
\end{equation}
The \index{Laplace transform}Laplace transformation yields 
\begin{equation}
Z_{\rm tot}(s)=\frac{R_{\rm p}}{1+s \frac{Q_{\rm p}}{\omega_{\rm res}}+\frac{Q_{\rm p} \omega_{\rm res}}{s}}
=\frac{R_{\rm p} \frac{\omega_{\rm res}}{Q_{\rm p}} s}{s \frac{\omega_{\rm res}}{Q_{\rm p}}+s^2+\omega_{\rm res}^2},
\label{glimpedancelaplace1}
\end{equation}
which may be found in literature (see Refs.~\cite{Pedersen1975,Boussard1997}) in the form 
\[
Z_{\rm tot}(s)=\frac{2 R_{\rm p} \sigma \; s}{s^2+2 \sigma s +\omega_{\rm res}^2}
\]
if $\sigma=\frac{\omega_{\rm res}}{2 Q_{\rm p}}$ is defined. 
As shown in Ref.~\cite{Klingbeil2015},  Eq.~(\ref{glimpedance1}) leads to the  
$3 \;\rm dB$ \index{Bandwidth!of cavity}bandwidth of the cavity provided that $R_{\rm p}$ and $Q_{\rm p}$ are frequency-independent: 
\[
\rahmen{
Q_{\rm p}=\frac{\omega_{\rm res}}{\Delta \omega_{\rm 3dB}}=\frac{f_{\rm res}}{\Delta f_{\rm 3dB}}
}.
\]

\index{Q factor|)}
\index{Resonant frequency|)}

\section{Length of the Cavity}

In the previous sections, we assumed that the ring cores can be 
regarded as lumped-element inductors and resistors. This is of course only 
true if the cavity is short in comparison with the wavelength. 

\index{Ring core|(}

As an alternative to the transformer model introduced above, one may therefore 
use a coaxial transmission line model \cite{Kanazawa2006}. For example, the section of the cavity
that is located on the left side of the~ceramic gap in Fig.~\ref{cavity} 
may be interpreted as 
a coaxial line that is homogeneous in longitudinal direction
and that has a short-circuit at the left end. 
The cross section consists of the magnetic material of the ring cores,  
air between the ring cores and the beam pipe, and air between the ring cores 
and the~cavity housing.  
This is of course an idealization since cooling disks, conductors and other 
air regions are neglected.  
Taking the SIS18 cavity at GSI as an example, the ring cores have 
$\mu_{\rm r}=28$ at an operating frequency of $2.5 \; \rm MHz$.
They have a relative dielectric constant of $10...15$, 
but this is reduced 
to an effective value of $\varepsilon_{\rm r,eff}=1.8$ since the ring cores do not 
fill the full cavity cross section.   
These values lead   
to a~wavelength of $\lambda=16.9 \; \rm m$. Since 64 ring cores with a thickness of 
$25 \; \rm mm$ are used, the effective length of the magnetic material is 
$1.6 \; \rm m=0.095 \;\lambda$ (which corresponds to a phase of $34^\circ$). 
In this case, the~transmission line model  
leads to deviations of less than $10 \; \%$ with respect to the lumped-element model.

The transmission line model makes it understandable why the MA- or ferrite-loaded cavity is sometimes 
referred to as a \textbf{shortened quarter-wavelength resonator}. 

Of course, one may also use more detailed models where subsections of the 
cavity are modeled as lumped elements. In this case, computer simulations 
can be performed to calculate the overall impedance. 
In case one is interested in resonances which may occur at higher operating 
frequencies, one should perform full electromagnetic simulations. 

In any case, one should always remember that some parameters are difficult to 
determine, especially the permeability of the ring core material under  
different operating conditions. This uncertainty may lead to larger errors 
than simplifications of the model. Measurements of full-size ring cores 
in the~requested operating range are inevitable when a new cavity is developed \cite{Klingbeil2020}. 
Also parameter tolerances due to the manufacturing process have to be taken 
into account. 

\index{Ring core|)}

In general, one should note that the total length and the dimensions of 
the cross-section of the~ferrite or MA cavity are not 
determined by the wavelength as for a conventional RF cavity. For example, 
the SIS18 ferrite cavity has a length of $3 \; \rm m$ flange-to-flange although only 
$1.6 \; \rm m$ are filled with magnetic material. 
This provides space for the ceramic gap, the cooling disks and further 
devices like the bias current bars. In order to avoid resonances at higher 
frequencies of the operating range, one should not waste too much space, but there is no exact  
size of the cavity housing that results from the electromagnetic analysis. 

\section{Differential Equation and Cavity Filling Time}

\label{cavityfillingtime}

\index{Impedance!cavity|see{Cavity, impedance}}
\index{Time constant!cavity|see{Cavity, time constant}}
\index{Cavity!time constant|(}

The equivalent circuit shown in Fig.~\ref{esbrlc2} was derived in the 
frequency domain. As long as no parasitic resonances occur, this equivalent 
circuit may be generalized. Therefore, we may also analyze it in the~time 
domain (allowing only slow tuning changes of $L_{\rm p}$ and $R_{\rm p}$ with time) leading to
\begin{align}
\rahmen{
\ddot V_{\rm gap}+\frac{2}{\tau} \; \dot V_{\rm gap}+\omega_{\rm res}^2 V_{\rm gap}=
\frac{1}{C} \tdq{}{t}(I_{\rm gen}-I_{\rm beam})
}.
\label{gldglvgap1}
\end{align}
Here we used the definition  
\[
\rahmen{
\tau=2 R_{\rm p} C
}.
\]
The product $R_{\rm p} C$ is also present in the expression for the quality factor:
\[
Q_{\rm p}=R_{\rm p} \sqrt{\frac{C}{L_{\rm p}}}=\frac{R_{\rm p} C}{\sqrt{L_{\rm p} C}}=\frac{1}{2} \tau \omega_{\rm res}
\mbox{\quad}
\Rightarrow
\rahmen{
\tau=\frac{2 Q_{\rm p}}{\omega_{\rm res}}=\frac{Q_{\rm p}}{\pi f_{\rm res}}
}
\]
Under the assumption $\omega_{\rm res}>\frac{1}{\tau}$ ($Q_{\rm p}>\frac{1}{2}$),
the ansatz $V_{\rm gap}=V_0 e^{\alpha t}$ (with a complex constant $\alpha$)
for the~homogeneous solution 
of Eq.~(\ref{gldglvgap1}) actually leads to 
$\alpha=-\frac{1}{\tau} \pm j \omega_{\rm d}$
with the exponential decay time $\tau$ and the oscillation frequency 
\[
\omega_{\rm d}=\omega_{\rm res} \sqrt{1-\frac{1}{(\tau \omega_{\rm res})^2}}
=\omega_{\rm res} \sqrt{1-\frac{1}{4 Q_{\rm p}^2}}.
\]
This yields $\omega_{\rm d}\approx \omega_{\rm res}$ even for 
moderately high $Q_{\rm p}>2$ (error less than $4 \; \%$). 
Sometimes, the resonant frequency $\omega_{\rm res}$ is
called the \index{Undamped natural frequency}\textbf{undamped natural frequency} whereas 
$\omega_{\rm d}$ is called the \index{Damped natural frequency}\textbf{damped natural frequency}. 

\index{Natural frequency!damped|see{Damped natural frequency}}
\index{Natural frequency!undamped|see{Undamped natural frequency}}

The time $\tau$ is the time constant for the cavity which also determines the
\index{Cavity!filling time}\textbf{cavity filling time}. 
Furthermore, the time constant $\tau$ is relevant for amplitude and  
phase jumps of the cavity (see Ref.~\cite{Papureanu1993}).

Sometimes, especially in the LINAC community, the cavity filling time is defined 
as $Q_{\rm p}/\omega_{\rm res}$ (one half of our definition, Ref.~\cite{Gerigk2011})
in order to specify the energy decay instead of the field strength or voltage decay.

\index{Cavity!impedance|)}
\index{Cavity!time constant|)}
\index{Filling time!cavity|see{Cavity, filling time}}

\section{Power Amplifier}

\index{Q factor!loaded|see{Loaded Q factor}}
\index{Q factor!unloaded|see{Unloaded Q factor}}
\index{RF power amplifier|see{Power amplifier}}
\index{Amplifier!power|see{Power amplifier}}

\index{Loaded Q factor|(}
\index{Unloaded Q factor|(}
\index{Power amplifier|(}

Up to now, the Q factor of the cavity was called $Q_{\rm p}$. What we did not mention
so far is that the Q factor of the cavity itself is the 
so-called \textbf{'unloaded Q factor'}. From now on, this unloaded Q factor 
will be denoted as $Q_{\rm p,0}$. In accordance with this, the parallel resistor
will be denoted as $R_{\rm p,0}$. The reason for this is the following:
An RF power amplifier that feeds the cavity may often be represented by a 
voltage-controlled current source (e.g.~in the case of a tetrode amplifier). 
The impedance of this current source will 
be connected in parallel to the equivalent circuit of the cavity thereby reducing the 
ohmic part $R_{\rm p}$ according to $R_{\rm p}=R_{\rm p,0}||R_{\rm gen}$
(see Fig.~\ref{gen_cav}, Section~\ref{resonant_frequency_control}). 
Therefore, the \textbf{loaded Q factor} $Q_{\rm p}$ will usually be reduced in comparison
with the unloaded Q factor $Q_{\rm p,0}$ provided that $C_{\rm gen}$ is not too large. 

The formulas that were derived for the parallel equivalent circuit are valid 
for both cases, the cavity alone and the combination of cavity and amplifier.
This is the reason why they were based on $R_{\rm p}$.

\index{Loaded Q factor|)}
\index{Unloaded Q factor|)}

It has to be emphasized that for MA- or ferrite-loaded cavities $50~\Omega$ 
impedance matching is not necessarily 
used in general. The cavity impedance is usually in the order of a few hundred 
ohms or a~few kilo-ohms. Therefore, it is often more suitable to directly 
connect the tetrode amplifier to the~cavity.   
Impedance matching is not mandatory 
if the amplifier is located close to the~cavity.
Short cables have to be used since they contribute to the overall impedance. 
Cavity and RF power amplifier must be considered as one unit; they cannot be
developed separately because their impedance curves influence each other. 

\index{Power amplifier|)}

\section{Cooling}
\label{cooling}

\index{Cooling|(}

Both, the power amplifier and the ring cores need active cooling. 
Depending on the operating conditions (e.g.~CW or pulsed operation), 
\textbf{forced air cooling} may be sufficient for the ring cores or \textbf{water cooling} may be required. 
Avoiding direct contact of the ferrite ring cores with water,
\textbf{cooling disks} in-between the cores may be used. In this case, one has
to make sure that the thermal contact between cooling disks and ferrites is 
good. For MA ring cores, direct water cooling \cite{Ohmori2013} or 
direct oil cooling \cite{Huelsmann2018} has also been used. 

\index{Cooling|)}

\section{Cavity Tuning}

\index{Cavity!tuning|see{Tuning}}
\index{Tuning|(}

We already mentioned in Section \ref{ferrites} that a DC bias current may 
be used to decrease $\mu_\Delta$ which results in a higher resonant frequency. 
This is one possible way to realize \textbf{cavity tuning}. Strictly speaking, 
one applies a quasi-DC bias current since the resonant frequency must be 
modified during a synchrotron machine cycle if it shall be equal to the 
variable RF frequency. Such a tuning of the resonant frequency $f_{\rm res}$ to the
RF frequency $f_{\rm RF}$ is usually desirable since the large $Z_{\rm tot}$ allows
to generate large voltages with moderate RF power consumption.   

Sometimes, the operating frequency range is small enough in comparison with the
bandwidth of the cavity that no tuning is required. This is often the case for MA cavities with sufficiently low Q factors (see Ref.~\cite{Huelsmann2018}). 

If tuning is required, one has at least two possibilities to realize it, 
bias current tuning or capacitive tuning. 
The latter may be realized by a variable capacitor 
(see Refs.~\cite{Morvillo2003,Huelsmann2004}) 
whose capacitance may be varied 
by a stepping motor. This mechanical adjustment, however, is only possible 
if the resonant frequency is not changed from machine cycle to machine cycle 
or even within one machine cycle. 

\index{Biasing!longitudinal|see{Parallel biasing}}
\index{Biasing!transverse|see{Perpendicular biasing}}
\index{Longitudinal biasing|see{Parallel biasing}}
\index{Transverse biasing|see{Perpendicular biasing}}

\index{Biasing!parallel|see{Parallel biasing}}
\index{Biasing!perpendicular|see{Perpendicular biasing}}
\index{Biasing|(}
\index{Parallel biasing|(}
\index{Perpendicular biasing|(}

In the case of bias current tuning for ferrite ring cores, one has two different choices, namely 
\textbf{perpendicular biasing} (also denoted as \textbf{transverse biasing}) and 
\textbf{parallel biasing} (also denoted as \textbf{longitudinal biasing}). 
Also a mixture of both is possible~\cite{Vollinger2013}.
The terms parallel and perpendicular 
refer to the orientation of the DC field $H_{\rm bias}$ in comparison with the 
RF field $H$. 

Parallel biasing is simple to realize. One adds bias current 
loops which may in principle be located in the same way as the 
inductive \index{Coupling loop}coupling loop shown in Fig.~\ref{cavity}. 
If only a few loops are present,  
current bars with large cross sections are needed to withstand the bias current 
of several hundred amps. The~required DC current may of course be reduced if the 
number $N_{\rm bias}$ of loops is increased accordingly (keeping the ampere-turns constant). 
This increase of the number of bias current windings may be limited by 
resonances or by the inductance that is allowed for the bias current circuit (to realize the required tuning speed). 
On the other hand, a minimum number of current loops is usually 
applied to guarantee a~certain amount of symmetry which leads to a more 
homogeneous flux in the ring cores. 

\index{Q loss effect|(}
\index{High loss effect|(}
\index{Anomalous loss effect|(}

Perpendicular biasing is more complicated to realize; it requires more 
space between the ring-cores, and the permeability range is smaller than for 
parallel biasing. The main reason for using perpendicular biasing is that 
lower losses can be reached (see Ref.~\cite{Smythe1985}). 
One can also 
avoid the so-called \textbf{Q-loss effect} or \textbf{high loss effect}. 
The Q-loss effect often occurs in ferrites when 
parallel biasing is applied and if the bias current is constant or 
varying only slowly. After a few milliseconds, one observes that the induced 
voltage breaks down by a certain amount even though the same amount of RF 
power is still applied (see Refs.~\cite{Griffin1979,Kaspar2004}). 
For perpendicular biasing, the Q-loss effect was not 
observed. 
The~Q-loss effect is not fully understood yet. However,
there are strong indications that it may be caused by mechanical resonances of 
the ferrite ring cores induced by magnetostriction effects \cite{Koenig2008}. 
It was possible to suppress the Q-loss effect by mechanical
damping. For example, in some types of ferrite cavities, the~ring cores are 
embedded in a sealing compound \cite{Arbuzov2004} which should damp mechanical oscillations.   
Not only the~Q-loss effect but also further anomalous loss effects have been
observed \cite{Griffin1979}.

\index{Q loss effect|)}
\index{High loss effect|)}
\index{Anomalous loss effect|)}

\index{Biasing|)}
\index{Parallel biasing|)}
\index{Perpendicular biasing|)}

When the influence of biasing is described, one usually defines an average 
bias field $H_{\rm bias}$ for the~ring cores. For this purpose, one may use the 
magnetic field 
\[
H_{\rm bias}= \frac{N_{\rm bias} I_{\rm bias}}{2 \pi \bar r}
\]
located at the mean radius $\bar r=\sqrt{r_{\rm i} r_{\rm o}}$.
Of course, this choice is somewhat arbitrary from a theoretical point of view,
but it is based on practical experience. 

A combination of bias current tuning and capacitive tuning has also been 
applied to extend the~frequency range \cite{Pei1993}.

\section{Further Complications for Ferrite Ring Cores}

We already mentioned that the effective differential permeability depends 
on the hysteresis behavior of the material, i.e.~on the history of bias and
RF currents. It was also mentioned that, due to the spatial dimensions of the 
cavity, we have to deal with ranges 
between lumped-element circuits and distributed elements. 
The anomalous loss effects are a third complication. 
There are further 
points which make the~situation even more complicated in practice: 
\begin{itemize}
\item If no biasing is applied, the maximum of the magnetic field is present at 
the inner radius $r_{\rm i}$. One has to make sure that the maximum ratings of the
material are not exceeded. 
\item Bias currents lead to an $\rho^{-1}$ dependency of the induced magnetic 
field $H_{\rm bias}$. Therefore, biasing is more effective in the inner parts
of the ring cores than in the outer parts resulting in a $\mu_\Delta$ which 
increases with $\rho$. According to Eq.~(\ref{glbfield1}), this will modify the 
$\rho^{-1}$ dependency of the magnetic RF field. As a result, the dependence on 
$\rho$ may be much weaker than without bias field. 
\item The permeability does not only depend on the frequency, on the magnetic
RF field and on the~biasing. It is also temperature-dependent. 
\item Depending on the thickness and the conductivity of the ring cores, 
on the material losses 
and on the~operating frequency, the magnetic field may decay from the surface 
to the inner regions reducing the effective volume. 
\item At higher operating frequencies with strong bias currents, the 
differential permeability will be rather low. This means that the magnetic flux 
will not perfectly be guided by the ring cores anymore. The fringe fields in the
air regions will be more important, and resonances may occur. 
\end{itemize}

\section{Resonant Frequency Control}
\label{resonant_frequency_control}

\index{Cavity!equivalent circuit|(}

\begin{figure}[htb]
\begin{center}
\begin{minipage}[t]{10.2cm}
%\epsfxsize=12cm
%\epsffile{\bilderpfad/bilder/ESB_gen_cav1.eps}
\includegraphics[width=10.2cm]{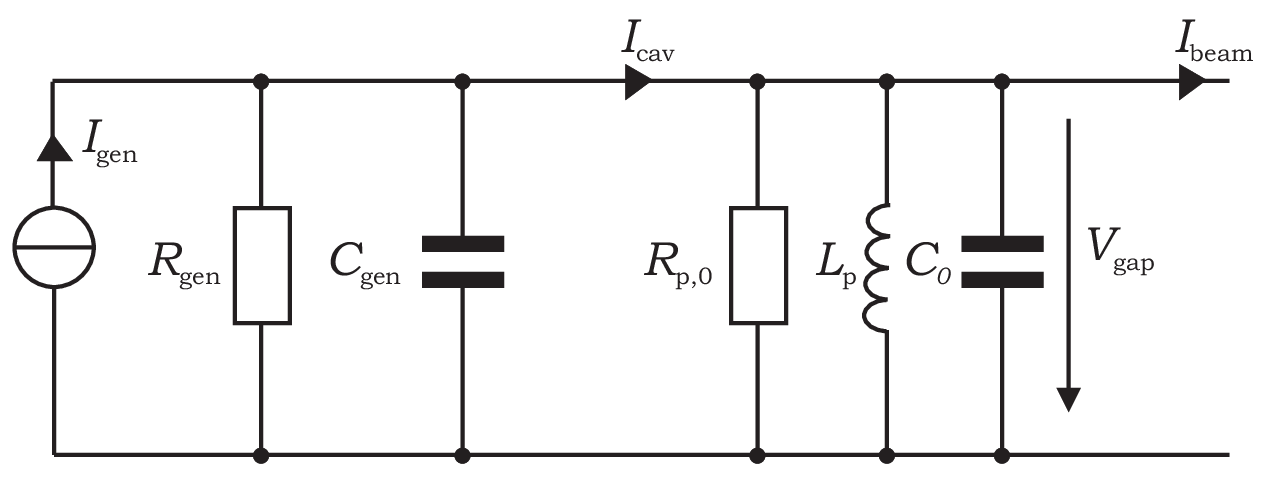}
\end{minipage}
\end{center}
\caption[bild]{Equivalent circuit of RF generator and cavity.}
\label{gen_cav}
\end{figure}

\index{Control grid|(}

A method that has traditionally been applied to decide whether a ferrite-loaded cavity is 
at resonance is the measurement of the 
phase of the gap voltage and 
of the phase of the control grid voltage (in case of a tetrode power amplifier).

At first sight it seems to be clear that the cavity is at resonance 
if and only if the cavity impedance is purely resistive, 
i.e.~if the inductance of the lumped element (parallel) circuit
exactly compensates the~capacitance (see Fig.~\ref{esbrlc2}). 
Therefore, it is obvious that the generator current and the gap voltage 
must be in-phase for the cavity operating at resonance. 

In order to analyze this fact in detail, however, one should be aware that a tube amplifier also contributes 
to the overall cavity impedance. Not only the ohmic output impedance will contribute to the overall impedance 
of the cavity, but also the capacitance of the tetrode (internal) and its circuitry (and also some inductances).  
As shown in Fig.~\ref{gen_cav} we may use a model~\cite{Hartel2011} in which the tetrode power amplifier is represented as 
a voltage-controlled current source with an internal resistor $R_{\rm gen}$ in parallel. The capacitance of the power amplifier
is represented by a capacitor $C_{\rm gen}$ (if necessary, $C_{\rm gen}$ may be frequency-dependent to include the
effect of parasitic inductances). The cavity without the power amplifier is shown on the right side of the 
circuit. It consists of $R_{\rm p,0}$, $L_{\rm p}$, and $C_0$. Now it becomes obvious that the resonant frequency
of the overall system is 
\[
\omega_{\rm res}=\frac{1}{\sqrt{L_{\rm p} (C_0+C_{\rm gen})}}.
\]
If the whole cavity system is tuned to resonance, it is not the current $I_{\rm cav}$ which is in-phase with the~gap voltage $V_{\rm gap}$,
but the current $I_{\rm gen}$. This current $I_{\rm gen}$ cannot be measured, however, since it may be regarded
as an internal current of the tetrode. Fortunately, the control grid voltage $V_{\rm g1}$ of the tetrode, however,
usually is in phase 
with the internal current $I_{\rm gen}$. Therefore, 
a \index{Resonant frequency control}\textbf{resonant frequency control} loop may compare the phase of
$V_{\rm g1}$ with the phase of $V_{\rm gap}$. If both are in phase (or $180^\circ$ out of phase, depending on the
chosen orientation of the voltages), the whole cavity system consisting of generator and cavity will be at resonance. 
One typically uses a \index{Phase detector}\textbf{phase detector} that provides this phase difference between $V_{\rm g1}$ and $V_{\rm gap}$.
This output signal is then used by a closed-loop controller 
to modify the bias current of the ferrite cavity
such that $I_{\rm gen}$ and $V_{\rm g1}$ are in phase with $V_{\rm gap}$ as desired.

\index{Control grid|)}
\index{Detector!phase|see{Phase detector}}
\index{Control!resonant frequency|see{Resonant frequency control}}

It has to be emphasized that the equivalent circuit in Fig.~\ref{esbrlc2}
and the related formulas are 
still valid. We just have to interpret the circuit in a slightly different way. 
If we compare it with Fig.~\ref{gen_cav} it becomes obvious that 
$C=C_{\rm gen}+C_0$ is the total capacitance of generator and cavity,
and $R_{\rm p}=R_{\rm p,0}||R_{\rm gen}=\frac{R_{\rm p,0} R_{\rm gen}}{R_{\rm p,0}+R_{\rm gen}}$ also
includes both contributions.  

\index{Cavity!equivalent circuit|)}

Sometimes it is desired not to operate the cavity at resonance but slightly off-resonance. 
In this case, one may choose a target value that differs from zero (or $180^\circ$, respectively) for the phase difference.
 
Completely detuning the cavity may be a choice to de-activate the cavity without having a high beam impedance. 
Then, of course, no gap voltage is produced so that the closed-loop control system will not work. However, 
one may modify the bias current in an open-loop mode in this case. 

\index{Tuning|)}

\section{Cavity Configurations}

A comparison of different types of ferrite- or MA-loaded cavities can be found in Refs.~\cite{Susini1988, Gardner1991, Schnase2000}. 
We just summarize a few aspects here that lead to different solutions. 

\begin{itemize}
\item Instead of using only one stack of ring cores and only one 
ceramic gap as shown in Fig.~\ref{cavity}, one may also use more sections 
with ring cores (e.g.~one gap with half the ring cores on the left side and the other 
half on the right side of the gap --- for reasons of symmetry) or more gaps. 
Sometimes, the ceramic gaps belong to different independent cavity cells 
which may be coupled by copper bars (e.g.~by connecting them in parallel~\cite{Huelsmann2018}).
Connections of this type must be short to allow operation at high frequencies. 
\item One configuration that is often used is a cavity consisting of only 
one ceramic gap and two ferrite stacks on both sides. Figure-of-eight windings (for a bias current)
surround these two ferrite stacks (see Ref.~\cite{Krusche1998}). 
With respect to the magnetic RF field, this 
leads to the same magnetic flux in both stacks. In this way, an RF power amplifier 
that feeds only one of the two cavity halves will indirectly supply the other 
cavity half as well. This corresponds to a 1:2 transformation ratio. Hence, the
beam will see four times the impedance compared with the amplifier load. 
Therefore, the same RF input power will lead to higher gap voltages (but also 
to a higher beam impedance). The transformation law may be derived by an analysis
that is similar to the one in Section \ref{analysis}.
\item Instead of the inductive coupling shown in Fig.~\ref{cavity}, one may 
also use capacitive coupling if the~power amplifier is connected to the gap 
via capacitors. If a tetrode power amplifier is used, one still has to 
provide it with a high anode voltage. Therefore, an external inductor 
\index{Choke coil}(choke coil) is 
necessary which allows the DC anode current but which blocks the RF current from 
the DC power supply. Often a combination of capacitive and inductive 
coupling is used (e.g.~to influence parasitic resonances). 
The coupling elements will contribute to the equivalent circuit. 
\item Another possibility is inductive coupling of individual ring cores. 
This leads to lower impedances which ideally allow a $50~\Omega$ impedance 
matching to a standard solid-state RF power amplifier (see Ref.~\cite{Dey1995}).  
\item In case a small relative tuning range is required, it is not necessary to use 
biasing for the ferrite ring cores inside the cavity. One may use external 
tuners (see Refs.~\cite{Hutcheon1987,Friedrichs1987}) which can be connected 
to the gap. For external tuners, both parallel and 
perpendicular biasing may be applied \cite{Poirier1993}.
\end{itemize}

No general strategy can be defined how a new cavity is designed. Many 
compromises have to be found. 
A certain minimum capacitance is given by the gap capacitance, the power amplifier capacitance, and 
parasitic capacitances. In order to reach the upper limit of the frequency range, 
a certain minimum inductance has to be realized. 
If biasing is used, this minimum inductance must be reached using the~maximum bias current but the effective permeability should still be high enough
to reduce stray fields. Also the lower frequency limit should be reachable with 
a minimum but non-zero bias current. There is a~maximum RF field 
$B_{\rm RF,max}$ (about~$10 \dots 20 \; \rm mT$ for ferrites) which should not be exceeded 
for the ring cores. 
This imposes a lower limit for the number of ring cores. The required tuning 
range in combination with the overall capacitance will also restrict the number
of ring cores. As mentioned above, the amplifier design should be 
taken into account from the very beginning, especially with respect to the~impedance. The maximum \index{Beam impedance}\textbf{beam impedance} that is 
tolerable is defined by
beam dynamics considerations. This impedance budget also defines the power 
that is required. If more ring cores can be used, the impedance of the cavity 
will increase, and the power loss will decrease for a given gap voltage.

\section{The GSI Ferrite Cavities in SIS18}
\label{SIS18_ferrite_cav}

\begin{figure}[htb]
\begin{minipage}[t]{7.5cm}
%\includegraphics[width=12cm]{\bilderpfad/bilder/SIS18_cavity}
%\epsfxsize=12cm
%\epsffile{\bilderpfad/bilder/SIS18_cavity.eps}
\includegraphics[width=7.5cm]{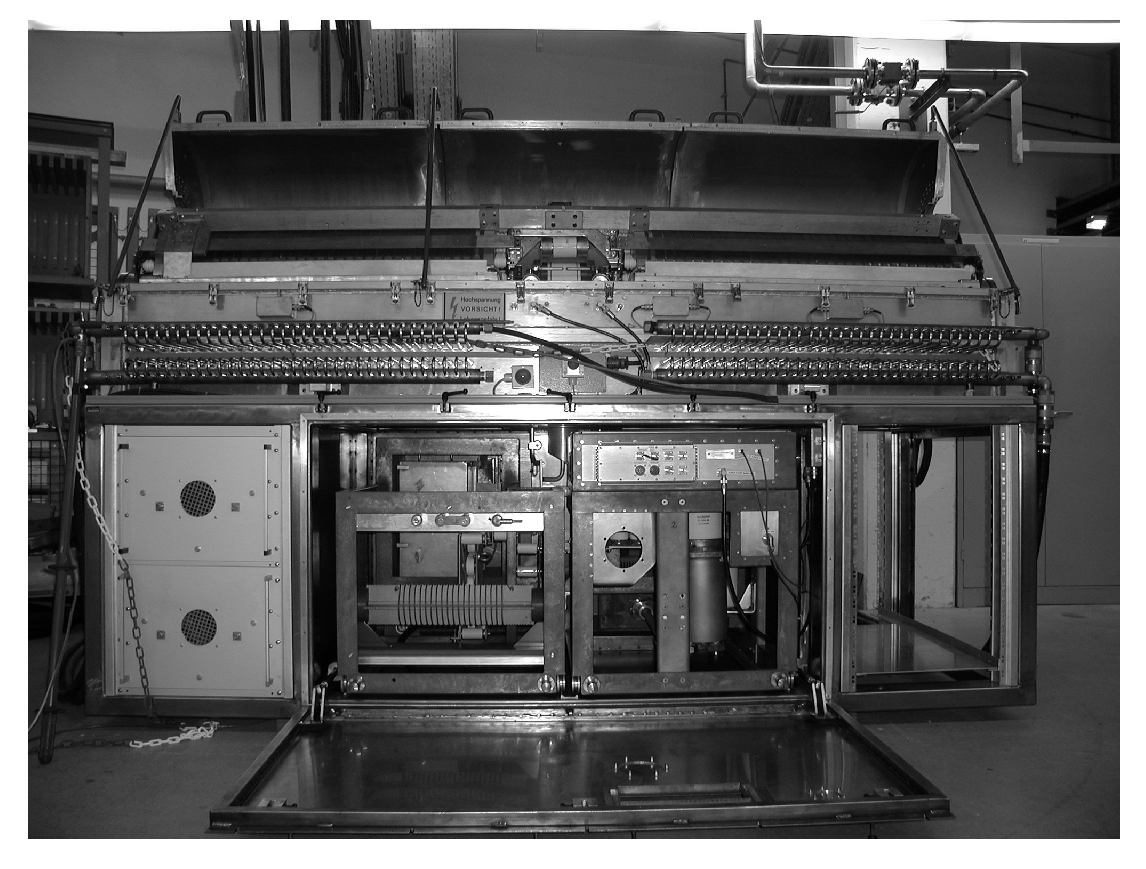}
\caption[bild]{SIS18 ferrite cavity. 
Photography: GSI Helmholtzzentrum f\"ur Schwerionenforschung GmbH, T.~Winnefeld.}
\label{sis18_cavity}
\end{minipage}
\hfill 
\begin{minipage}[t]{7.5cm}
%\includegraphics[width=12cm]{\bilderpfad/bilder/Gap}
%\epsfxsize=12cm
%\epsffile{\bilderpfad/bilder/Gap.eps}
\includegraphics[width=7.5cm]{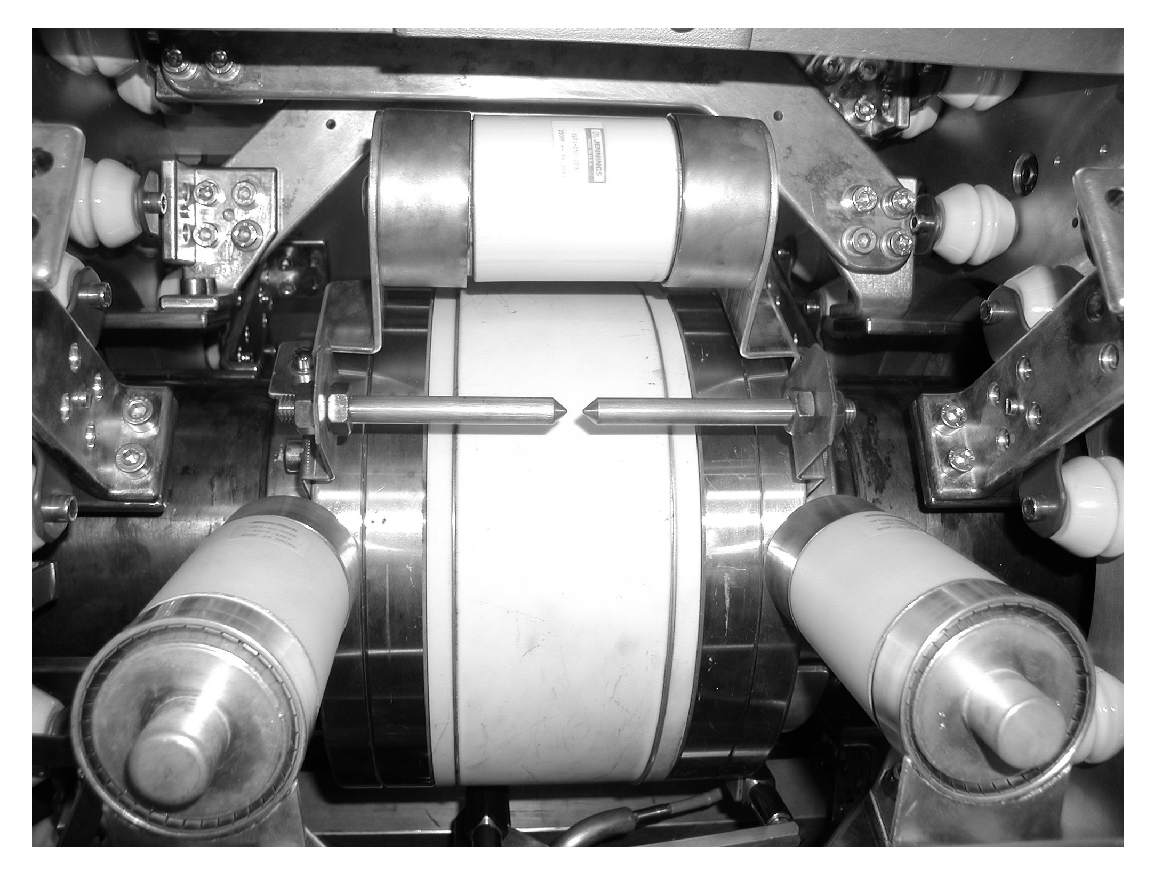}
\caption[bild]{Gap area of the SIS18 ferrite cavity.
Photography: GSI Helmholtzzentrum f\"ur Schwerionenforschung GmbH, T.~Winnefeld.}
\label{gap_sis18_cavity}
\end{minipage}
\end{figure}

\begin{table}[h!]
\centering
\caption{Equivalent circuit parameters for SIS18 ferrite cavities (without resistance of tetrode amplifiers).}
\begin{tabular}{l|l|l|l} \hline \hline
\textbf{Resonant frequency} $\bm{f_{\rm res}}$ & $620 \; \rm kHz$ & $2.5 \; \rm MHz$ & $5 \; \rm MHz$\\ \hline
\textbf{Relative permeability} $\bm{\mu'_{\rm p,r}}$ & $450$ & $28$ & $7$\\ \hline
\textbf{Magnetic bias field $\bm{H_{\rm bias}}$ at mean radius} & $25 \; \rm A/m$ & $700 \; \rm A/m$ & $2750 \; \rm A/m$ \\ \hline 
\textbf{Bias current $\bm{I_{\rm bias}}$} & $4.8 \; \rm A$ & $135 \; \rm A$ & $528 \; \rm A$ \\ \hline 
\textbf{$\bm{\mu'_{\rm p,r} Q f}$ product} & $4.2 \cdot 10^9 \; \rm s^{-1}$ & $3.7 \cdot 10^9 \; \rm s^{-1}$ & $3.3 \cdot 10^9 \; \rm s^{-1}$ \\ \hline
\textbf{Q-factor $\bm{Q_{\rm p,0}}$} & $15$ & $53$ & $94$ \\ \hline 
\textbf{$\bm{L_{\rm s}}$} & $88.2~\rm \mu H$ & $5.49~\rm \mu H$ & $1.37~\rm \mu H$ \\ \hline
\textbf{$\bm{L_{\rm p}}$} & $88.5~\rm \mu H$ & $5.49~\rm \mu H$ & $1.37~\rm \mu H$ \\ \hline
$\bm{R_{\rm s,0}}$ & $22.8~\rm \Omega$ & $1.63~\rm \Omega$ & $0.46~\rm \Omega$ \\ \hline 
$\bm{R_{\rm p,0}}$ & $5200~\rm \Omega$ & $4600~\rm \Omega$ & $4100~\rm \Omega$ \\ \hline
\textbf{Cavity time constant} $\bm{\tau}$ & $7.7~\rm \mu s$ & $6.7~\rm \mu s$ & $6.0~\rm \mu s$ \\ \hline \hline
\end{tabular} 
\label{tabesbparam}
\end{table} 

As an example for a ferrite cavity, we summarize the main facts about GSI's 
SIS18 ferrite cavities (see Figs.~\ref{sis18_cavity} and \ref{gap_sis18_cavity}). 
Two identical ferrite cavities are located in the synchrotron SIS18. 

The material Ferroxcube FXC 8C12m is used for the ferrite ring cores.  
In total, $N=64$ ring cores are used per cavity. Each core has the following 
dimensions: 
\[
d_{\rm o}=2 \; r_{\rm o}=498 \; {\rm mm}\mbox{,\quad}
d_{\rm i}=2 \; r_{\rm i}=270 \; {\rm mm}\mbox{,\quad}
d_{\rm core}=25 \; {\rm mm} \mbox{,\quad}
\bar r=\sqrt{r_{\rm i} r_{\rm o}}=183 \; \rm mm
\]
For biasing, $N_{\rm bias}=6$ figure-of-eight copper windings are present. 
The total capacitance amounts to $C=740 \; \rm pF$
including the gap, the gap capacitors, the cooling disks, and other parasitic 
capacitances. The maximum voltage that is reached under normal operating 
conditions is $\hat V_{\rm gap}=16 \; \rm kV$.

Table \ref{tabesbparam} shows the main parameters for three different 
frequencies. All these values are consistent with the formulas presented 
in this article. It is obvious that both $\mu'_{\rm p,r} Q f$ and $R_{\rm p,0}$
do not vary strongly with 
frequency justifying the parallel equivalent circuit. 
This compensation effect was mentioned at the~end of Section \ref{material}.
 
All the parameters mentioned here refer to the beam side of the cavity. 
The cavity is driven by an~RF amplifier which is coupled to 
only one of two ferrite core stacks (consisting of 32 ring cores each). 
The~two ring core stacks are coupled by the bias windings. Therefore, a 
transformation ratio of 1:2 is present from amplifier to beam.
This means that the amplifier has to drive a load of about 
$R_{\rm p,0}/4 \approx 1.1~\rm k\Omega$. For a full amplitude 
of $\hat V_{\rm gap}=16 \; \rm kV$ at
$f=5\; \rm MHz$ the power loss in the cavity amounts to $31 \; \rm kW$.

The SIS18 cavity is supplied by a single-ended tetrode power amplifier using 
a combination of inductive and capacitive coupling. 

It has to be emphasized that the values in Table \ref{tabesbparam} do not 
contain the amplifier resistance. Depending on 
the \index{Operating point}operating point of the tetrode,
$R_{\rm p}$ will be reduced significantly in comparison with $R_{\rm p,0}$, 
and all related parameters vary accordingly.

\section{Further Practical Considerations}

\index{Gap voltage divider|(}
\index{Gap relay|(}
\index{Gap switch|(}

For measuring the gap voltage, one needs a \textbf{gap voltage divider} in order to 
decrease the high-voltage RF to a safer level. This can be done by capacitive 
voltage dividers. 
\textbf{Gap relays} are used to short-circuit the gap if the cavity is temporarily 
unused. This reduces the \index{Beam impedance}beam impedance which may be harmful for 
beam stability. 
If cycle-by-cycle switching is needed, semiconductor switches may be used 
as gap switches 
instead of vacuum relays. Another possibility to temporarily reduce the 
beam impedance is to detune the cavity.  

\index{Gap voltage divider|)}
\index{Gap relay|)}
\index{Gap switch|)}

The capacitance/impedance of the gap periphery devices must be considered 
when the overall capacitance $C$ and the other elements 
in the equivalent circuit are calculated. 
Also further parasitic elements may be present. 

On the one hand, the cavity dimensions should be as small as possible since
space in synchrotrons and storage rings is valuable and since undesired 
resonances must be avoided. On the other hand certain minimum distances have
to be kept in order to prevent high-voltage spark-overs. 
Due to \index{EMC}\textbf{EMC} (\textbf{electromagnetic compatibility}) reasons, 
RF seals are often used between conducting metal parts of
the~cavity housing to reduce electromagnetic emission. 

\index{Electromagnetic compatibility|see{EMC}}

In order to fulfill high vacuum requirements, it may be necessary to allow 
a \index{Bakeout}\textbf{bakeout} of the vacuum chamber. This can be realized by 
integrating a \index{Heating jacket}\textbf{heating jacket} that surrounds 
the beam pipe. It has to be guaranteed that the ring cores
are not damaged by heating and that safety distances 
(for RF purposes and high-voltage requirements) are kept.  

In case the cavity is used in a radiation environment, the radiation hardness
of all materials is an~important topic. 

\section{Magnetic Materials}

\index{Magnetic material|(}

A large variety of magnetic materials is available. 
Nickel-Zinc (NiZn) ferrites may be regarded as the~traditional standard material 
for ferrite-loaded cavities. At least the following material properties are of interest
for the material selection and may differ significantly for different types 
of material: 
\begin{itemize}
\item permeability, 
\item magnetic losses, 
\item saturation induction (typically $200 \dots 500 \; \rm mT$ for NiZn ferrites), 
\item maximum RF inductions (typically $10 \dots 20 \;\rm mT$ for NiZn ferrites),
\item relative dielectric constant (in the order of $10...15$ for NiZn ferrites but 
e.g.~very high for MnZn ferrites) and dielectric losses (usually negligible for typical
NiZn applications),
\item maximum operating temperature, thermal conductivity and temperature dependence in general,  
\item magnetostriction, 
\item specific resistance (very high for NiZn ferrites, very low for MnZn ferrites).
\end{itemize}
In order to determine the RF properties under realistic operating conditions
(large magnetic flux, biasing), thorough reproducible measurements in a
fixed test setup are inevitable~\cite{Klingbeil2020}. 

Amorphous and nanocrystalline magnetic alloy (MA) materials have been used 
to build very compact cavities that are based on similar principles as the 
classical ferrite cavities 
(see Refs.~\cite{Fougeron1993,Schnase2000,Saito2001,Huelsmann2004,Ohmori2005,Kanazawa2006,Stassen2008,Mei2012,
Ohmori2013,Wang2021}). 
These materials have very high saturation induction ($500 \dots 1600 \; \rm mT$) \cite{Coey}. 
They allow a~higher RF induction (provided that cooling is sufficient) and have a very high permeability. 
This means that a~smaller number of ring cores is needed for the same inductance. 
MA materials typically have lower Q factors in comparison with ferrite materials. 
Low Q factors have the advantage that frequency tuning is often not necessary (see Ref.~\cite{Huelsmann2018})
and that it is possible to generate signal forms 
including higher harmonics \cite{Fujieda1999,Ohmori2005b} instead of pure sine signals. 
MA cavities are especially of interest for pulsed operation at high field 
gradients~\cite{Huelsmann2004,Ohmori2005,Mei2012}. In case a low Q-factor
is not desired, it is also possible to increase it by cutting the MA ring cores to introduce air gaps \cite{Ohmori2005,Ohmori2013}. 
The manufacturing process of MA ring cores differs significantly from that of ferrite ring cores \cite{Coey}. For ferrites, a metal oxide powder is the basis, and sophisticated milling, pressing, sintering and machining/finishing processes are performed. This leads to a homogeneous ring core with a flat surface. In order to produce an MA tape/ribbon, melted metal is disposed on a cooled rotating wheel/drum by means of a nozzle (planar flow casting \cite{Coey}). The ring core is manufactured by winding the ribbon (including an insulation layer), performing elaborated annealing processes and coating/wrapping/impregnation. Therefore, MA cores are radially inhomogeneous, and the~surface is not intrinsically flat. These differences concerning the manufacturing processes are important for cooling considerations (see Section~\ref{cooling}) and also for the challenge to control the tolerances of the~ring core parameters listed above. 

Microwave garnet ferrites have been used at frequencies in the 
range $40...60 \;\rm MHz$ (or even wider) in connection with perpendicular biasing 
since they provide comparatively low losses 
(see Refs.~\cite{Earley1983,Kaspar1984,Schaffer1992,Eberhardt2016,Madrak2016}).   

\index{Magnetic material|)}

\index{Ferrite-loaded cavity|)}

\section*{Acknowledgements}

The author would like to thank all the current and former GSI colleagues with whom he discussed several RF cavity issues
during the past years, especially Priv.-Doz.~Dr.~habil.~Peter Hülsmann, Dr.~Hans Günter König, Dr.~Ulrich Laier, Michael Frey, 
and Dr.~Gerald Schreiber. He is also grateful to the former GSI members, especially
Dr.~Klaus Blasche, Dipl.-Phys.~Martin Emmerling, and Dr.~Klaus Kaspar for transferring their RF cavity
know-how to their successors. Last but not least, the author thanks Dr.~Rolf Stassen (FZ Jülich) for
reviewing earlier versions of the manuscript and Dr.~Ulrich Laier for his helpful comments that led to an improvement of this latest version.

It is impossible to provide a complete list of references. The following list cites only a few references regarding the most important aspects. Many other important publications exist.

\bibliographystyle{unsrtenglkb}

\bibliography{lit}

\end{document}